\documentclass[conference]{IEEEtran}
\IEEEoverridecommandlockouts
\usepackage{graphicx}
\usepackage{amsmath}
\usepackage{amsfonts}
\usepackage{mathrsfs}
\usepackage[OT1]{fontenc} 
\usepackage{enumerate}
\usepackage{amsthm}
\usepackage{tabularx}
\usepackage{arydshln}

\usepackage[colorlinks=true,linkcolor=blue,citecolor=blue]{hyperref}
\usepackage{mwe}
\usepackage{subfig}
\usepackage{float}
\usepackage{graphicx}
\usepackage{textcomp}
\usepackage{xcolor}
\usepackage{footnote}
\usepackage{makecell}
\usepackage{graphicx}
\usepackage{amsmath}
\usepackage{amsfonts}
\usepackage{mathrsfs}
\usepackage[OT1]{fontenc} 
\usepackage{enumerate}
\usepackage{amsthm}
\usepackage{tabularx}
\usepackage{arydshln}

\usepackage{url}
\usepackage{breakurl}

\usepackage{subfig}
\usepackage{textcomp}
\usepackage{xcolor}
\usepackage{footnote}
\usepackage{makecell}
\usepackage[draft]{todonotes}

\usepackage{algorithm}% http://ctan.org/pkg/algorithms
\usepackage{algpseudocode}% http://ctan.org/pkg/algorithmicx
\usepackage[inline]{enumitem}
\usepackage{mathtools}

\newcommand{\RNum}[1]{\uppercase\expandafter{\romannumeral #1\relax}}
\begin{document}
\title{Alternative Intersection Designs with Connected and Automated Vehicle}

%\author{\IEEEauthorblockN{Zijia Zhong \\ Earl E. Lee \RNum{2}\\ Mark Nejad }
%\IEEEauthorblockA{Department of Civil and Environmental Engineering\\
%University of Delaware\\
%Newark, DE, USA\\
%Email: \{zzhong, elee, nejad\}@udel.edu}
%}

\author{
\IEEEauthorblockN{ Zijia Zhong} %, \textit{member} IEEE
\IEEEauthorblockA{\textit{Department of Civil and Environmental Engineering} \\
\textit{University of Delaware}\\
Newark, DE, USA \\
zzhong@udel.edu}
\and
\IEEEauthorblockN{Earl E. Lee \RNum{2}}
\IEEEauthorblockA{\textit{Department of Civil and Environmental Engineering} \\
\textit{University of Delaware}\\
Newark, DE, USA \\
elee@udel.edu}
}

\maketitle
\begin{abstract}
Alternative intersection designs (AIDs) can improve the performance of an intersection by not only reducing the number of signal phases but also change the configuration of the conflicting points by re-routing traffic. However the AID studies have rarely been extended to Connected and Automated Vehicle (CAV) which is expected to revolutionize our transportation system. 
In this study, we investigate the potential benefits of CAV to two AIDs: the diverging diamond interchange (DDI) and the restricted crossing U-turn  intersection.  The potential enhancements of AID, CAV, and the combination of both are quantified via microscopic traffic simulation.  We found that CAV is able to positively contribute to the performance of an intersection. However, converting an existing conventional diamond interchange (CDI) to a diverging one is a more effective way according to the simulation results. DDI improves the throughput of a CDI by 950 vehicles per hour, a near 20\% improvement; whereas with full penetration of CAV, the throughput of a CDI is increased only by 300 vehicles per hour. A similar trend is observed in the average delay per vehicle as well. 
Furthermore, we assess the impact for the driver's confusion, a concern for deploying AIDs, on the traffic flow. According to the ANOVA test,  the negative impacts of driver's confusion are of statistical significance.

\end{abstract}


\section{Introduction}

Signalized intersections are major sources of traffic delay and collision within modern transportation systems. The measures to improve the operational efficiency of a signalized intersection can be grouped in to four categories: 
\begin{enumerate*}[label=\arabic*)]
\item Optimization of Signal Timing and Phase, 
\item Conversion to a grade-separated interchange, 
\item Reconfiguration to alternative intersection designs (AIDs), and 
\item Adaptation of connected and automated vehicle (CAV) technology.
\end{enumerate*}

The traditional approach via signal optimization is no longer able to considerably alleviate congestion at signalized intersections in saturated condition \cite{dhatrak2010performance}. Grade-separation tends to incur a significant amount of infrastructure investment, which is difficult to economically justify under most circumstances. AIDs have the potential in improving the efficiency and safety of an intersection by strategically eliminating or changing the nature of the intersection conflict points. 

\begin{figure} [h]
	\centering
	\frame{\includegraphics[width=0.95\columnwidth]{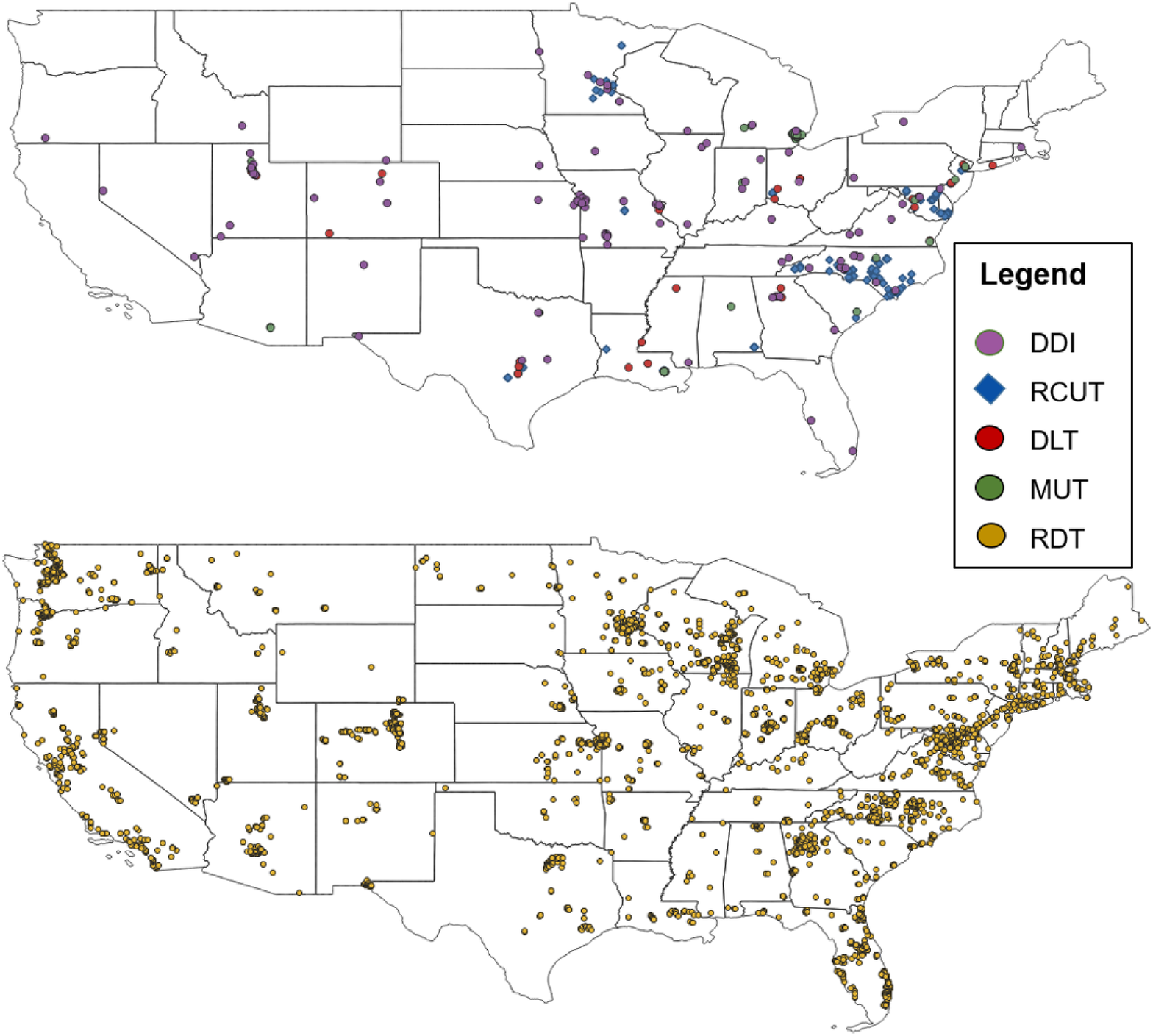}}   
	\caption{AID locations in contiguous U.S. (data source \cite{InstitueforTransportationResearchandEducation})} 
	\label{fig:AIDLocation}
\end{figure}

While the adoption of AIDs exhibits an increasing trend in the U.S. as displayed in Fig. \ref{fig:AIDLocation}, additional research for AID is still needed. The most common AIDs include the diverging diamond interchange (DDI), the median U-turn intersection (MUT), the displaced left-turn intersection (DLT), and roundabout (RDT).

\textcolor{black}{The evolutionary role of the CAV technology to mobility, safety, and driver convenience has been discussed extensively in the past decades. At the same time, the adaptation of AIDs has been growing steadily and their benefits have gained recognition. However, the joint benefits of implementing CAV and AID have been seldom discussed. The Volpe National Transportation Systems Center estimated that it may take 25-30 for CAVs to reach 95\% of market penetration (MPR), even with a federal mandatory installation of DSRC devices on new light vehicles in the U.S. \cite{volpe2008vehicle}. }

\textcolor{black}{
In light of the aforementioned lead time, hybrid solutions may be a logical step for solving the pressing transportation issues.}
In this paper, we evaluate the potential benefits brought by CAV, AID, and the combination of both. We also quantify the influence of the driver's confusion on a restricted crossing U-turn intersection (RCUT). Such driver's confusion that is caused by the unconventional geometry deign is expected to be eliminated by CAV.

%%%%%%%%%%%%%%%%%%%%%%%%%%%%%%%%%%%%%%%%%%%%%%%%%%%%%%%%%%%%%%%%%%%%%%%%%%%%%%%%%
\section{Related Work}
\label{sect:Literature}
\subsubsection{Effectiveness of AIDs}
The majority of the research demonstrated the superior performance of AIDs to their conventional counterparts under various volume scenarios,  for instance, heavy left-turn traffic, unbalanced split among intersection approaches, high overall volume, etc. Such scenarios can reveal the inadequacy of a conventional intersection. A diverging diamond interchange (DDI) outperform a conventional diamond interchange (CDI) under high traffic volume with left-turn demand exceeding 50\% of the total demand \cite{dhatrak2010performance}. When designed properly, the DDI can reduce 60\% of total intersection delay and 50\% of the total number of stops \cite{Chlewicki2011}.  A signal optimization model for DDI was developed in \cite{yang2014development}, in which the common cycle length and green split for the two up-stream crossover intersections were determined by taking into account the adjacent conventional intersections. 

The displaced left-turn (DLT) intersection is able to potentially reduce average intersection delays in most traffic demand scenarios. A before-and-after study for the DLT at Baton Rouge, LA showed that the reduction in total crashes and fatality were 24\% and 19\%, respectively. The simulation also demonstrated 20\% to 50\% increase in throughput compared to a conventional intersection \cite{hughes2010alternative}. The reduction for a median u-turn (MUT) intersection in total crashes ranges from 20\% to 50\%, as shown in the study conducted in \cite{scheuer1996evaluation,castronovo1995operational}. 

\subsubsection{Effectiveness of CAV}
A CAV-based application on real-world signalized intersection was studied using Vissim in \cite{Zhong2017a}. The start-up lost time was assumed to be zero owing to V2X communication. Addtionally, all the CAVs within a platoon operated synchronously upon the commencement of a green phase. Without changing the existing signal plan, the average stop delay was reduced by 17\% when the market penetration rate (MPR) of CAV reached 70\%. Le Vine et al. \cite{le2016automated} studied the queue discharging operation of CAVs with the assured-clear-distance-ahead principle by using a deterministic simulation model. On the contrary to \cite{Zhong2017a}, they observed only marginal improvement to intersection throughput due to the synchronous start-up movement.  However, they found that the processing time for a 10-vehicle queue did reduce by 25\% with full CAVs, compared to that for the human-driven vehicles (HVs) with the same amount of vehicles.

Realizing the potential long path to full vehicle automation, researchers also emphasized the possible cooperative scheme between CAVs and HVs by strategically consider the following HVs for intersection management \cite{le2016automated}. 
A  bi-level optimal intersection control algorithm was proposed in \cite{yang2016isolated}. The algorithm performed trajectory design for CAVs as well as the prediction for HVs based on real-time CAV data. The prediction of the trajectory of HVs was based on Newell’s car following model and the positional information of CAVs. The baseline used for comparison was an actuated signal control algorithm under a range of traffic demand between 1,000 and 2,000 vehicles per hour (vph). 

\subsubsection{Driver's Confusion}
Unfamiliar urban intersections pose high cognitive demand on drivers who are prone to make unexpected maneuvers, which include hesitation, abrupt stop, deviation from the planned path,  suddent aggressive maneuvers \cite{autey2013operational, sayed2006upstream}. The driver’s confusion was mentioned in most of the AID studies as a potential drawback.  As we observed from practices, the off-ramp right tuning movements from the freeway in DDIs are often signalized due to the safety concern for unfamiliar drivers who may misidentify traffic on the opposite side of the roadway passing through a DDI interchange \cite{chilukuri2011diverging}. Some believe that the reduction in delay and travel time would be discounted after accounting for driver’s confusion \cite{Reid2001}.  

A driving simulator provides a safe virtual environment for human subjects to experience a wide verity of scenario, including investigating the driver's confusion for AIDs.  In \cite{Bared2007}, 74 drivers within the Washington D.C. area were recruited for the experiment which aimed to investigate the wrong way violation, navigation errors, red-light violations, and driving speed through the DDI. In \cite{Claros2017}, Park found that wrong way crashes inside the crossroad between ramp terminals accounted for 4.8\% of the fatal and injury crashes occurring at the DDI.  
The CAV technology could be an excellent complement for the AIDs. The V2X connectivity is able to provide geometry information to help unfamiliar drivers to navigate through AIDs. Increasingly, the Automated Driver Assistant System could, when necessary, intervene with the erroneous movement as a result of the driver's confusion. Hence, the potential aid gained from CAV technology could improve the performance of AID by abating or even eliminating the concerns for the driver’s confusion. 
 
%%%%%%%%%%%%%%%%%%%%%%%%%%%%%%%%%%%%%%%%%%%%%%%%%%%%%%%%%%%%%%%%%%%%%%%%%%%%%%%%%

\section{Experiment}
\label{sect:framework}

The primary benefits for the introduction of CAV to AIDs are the enhanced driving performance due to automation and the connectivity with the signal controller. In other words, CAVs can closely follow their predecessors and have no driver's confusion for AIDs nor start-up lost time. We first demonstrate the improvement of AIDs with various penetration of CAVs for a diverging diamond interchange (DDI) and a restricted crossing U-turn intersection (RCUT). Then a proof-of-concept simulation for the impact of driver’s confusion is conducted.

%\todo[inline]{to describe the CAV behavioral model}
\textcolor{black}{Each CAV is assumed with SAE level 3 automation. The Enhanced Intelligent Driver Model (EIDM), developed by Kesting el al. \cite{Kesting2010} and expressed in (\ref{eq:eidm}), (\ref{eq: minDistCal}), and (\ref{eq: cahCal}), is adapted for longitudinal control), whereas the human drivers are responsible for the lateral control which is based on the Weidemann model \cite{Wiedemann1974, wiedemann1991modelling}. }

\begin{equation}
\ddot{x}=\begin{cases}
a[1-(\frac{\dot{x}}{\dot{x_{des}}})^{\delta }- (\frac{s^{*}(\dot{x}, \dot{x}_{lead})}{s_{0}})] & \\ \text{ if } x=  \ddot{x}_{IDM} \geq \ddot{x}_{CAH} \\ 
 (1-c)\ddot{x}_{IDM} + c[\ddot{x}_{CAH} + b \cdot tanh ( \frac{\ddot{x}_{IDM} - \ddot{x}_{CAH}}{b})] & \\\text{otherwise} 
\end{cases}
\label{eq:eidm}
\end{equation}
\begin{equation}
s^{*}(\dot{x}, \dot{x}_{lead}) = s_{0} + \dot{x}T + \frac{\dot{x}(\dot{x} - \dot{x}_{lead})}{2\sqrt{ab}} 
\label{eq: minDistCal}
\end{equation}
\begin{equation}
\ddot{x}_{CAH}=
\begin{cases}
\frac{\dot{x}^{2} \cdot \min(\ddot{x}_{lead}, \ddot{x})}{\dot{x}_{lead}^{2}-2x \cdot \min(\ddot{x}_{lead}, \ddot{x})} & \\
\dot{x}_{lead} (\dot{x} - \dot{x}_{lead}) \leq -2x \min(\ddot{x}_{lead}, \ddot{x})  \\
\min(\ddot{x}_{lead}, \ddot{x}) - \frac{(\dot{x}-\dot{x}_{lead})^{2} \Theta (\dot{x}- \dot{x}_{lead})}{2x}  &  \\ \text {otherwise}
\end{cases} 
\label{eq: cahCal}
\end{equation}

\textcolor{black}{where $a$ is the maximum acceleration; $b$ is the desired deceleration; $c$ is the coolness factor; $\delta$ is the free acceleration exponent; $\dot{x}$ is the current speed of the subject vehicle;  $\dot{x}_{des}$ is the desired speed,  $\dot{x}_{lead}$ is the speed of the lead vehicle; $s_{0}$ is the minimal distance; $\ddot{x}$ is the acceleration of the subject vehicle; $\ddot{x}_{lead}$ is the acceleration of the lead vehicle; $\ddot{x}_{IDM}$ is the acceleration calculated by the original IDM model \cite{Treiber2000}; $T$ is the desired time gap; and $\ddot{x}_{CAH}$ is the acceleration calculated by the CAH component; $\Theta$ is the Heaviside step function. The IDM parameters used are listed in TABLE \ref{table: parameters}}.

\begin{table}[!ht]
\centering
\caption{\textcolor{black}{CACC Vehicle Control Parameters}} 
\resizebox{\columnwidth}{!}
{
\begin{tabular}{cccccccccc}
\hline 
Parameter & $T$ & $s_{0}$ & $a$ & $b$ & $c$ & $\theta$ & $\dot{x}_{des}$  \\ \hline
value &  0.9 s & 1 $m$ & 2 $m/s^{2}$ & 2$m/s^{2}$ & 0.99 & 4 & 105 $km/h$  \\
\hline 
\end{tabular}
}
\label{table: parameters}
\end{table}

\textcolor{black}{
The benefits of AIDs and CAV are of complementary nature as exhibited in TABLE \ref{table:beneComp}. The primary benefit for CAV is the short following headway, which plays a crucial role in improving roadway capacity. Additionally, the elimination of start-up lost time (the time drivers takes to react and accelerate when a signal turns green from red) is also feasible owing to the vehicle-to-infrastructure (V2I) communication. The start-up lost time for HVs is set as 2 s. The effectiveness of the synchronized start has not been substantiated by previous research:  some reserach found significant benefits \cite{Zhong2017a}, while other did not \cite{le2016automated}.
Therefore, the first two benefits for CAV (close following headway and no start-up lost time) are implemented in the simulation. The simulation is conducted in two settings. First, we will evaluate the overall intersection performance. Then we shift the study focus on the region where driver's confusion could occur in order to assess its impact.
}
\begin{table}[h]
\centering
\caption{\textcolor{black}{Benefits of CAV and AID}}
\begin{tabular}{l|ll}
\hline
Benefit & AID & CAV \\ \hline
Intersection conflict point reduction & \checkmark  &  \\
Signal phase reduction & \checkmark & \\
Traffic movement streamlining & \checkmark  &\\
Close following headway & & \checkmark \\
Start-up lost time elimination & &\checkmark   \\
Synchronously discharge &  & \checkmark \\
Driver's confusion intervention & & \checkmark  \\
\hline
\end{tabular}
\label{table:beneComp}
\end{table}

\begin{figure}[h]
\begin{minipage}[h]{0.8\columnwidth}
\centering
\subfloat[DDI Network]{\includegraphics[scale=0.26]{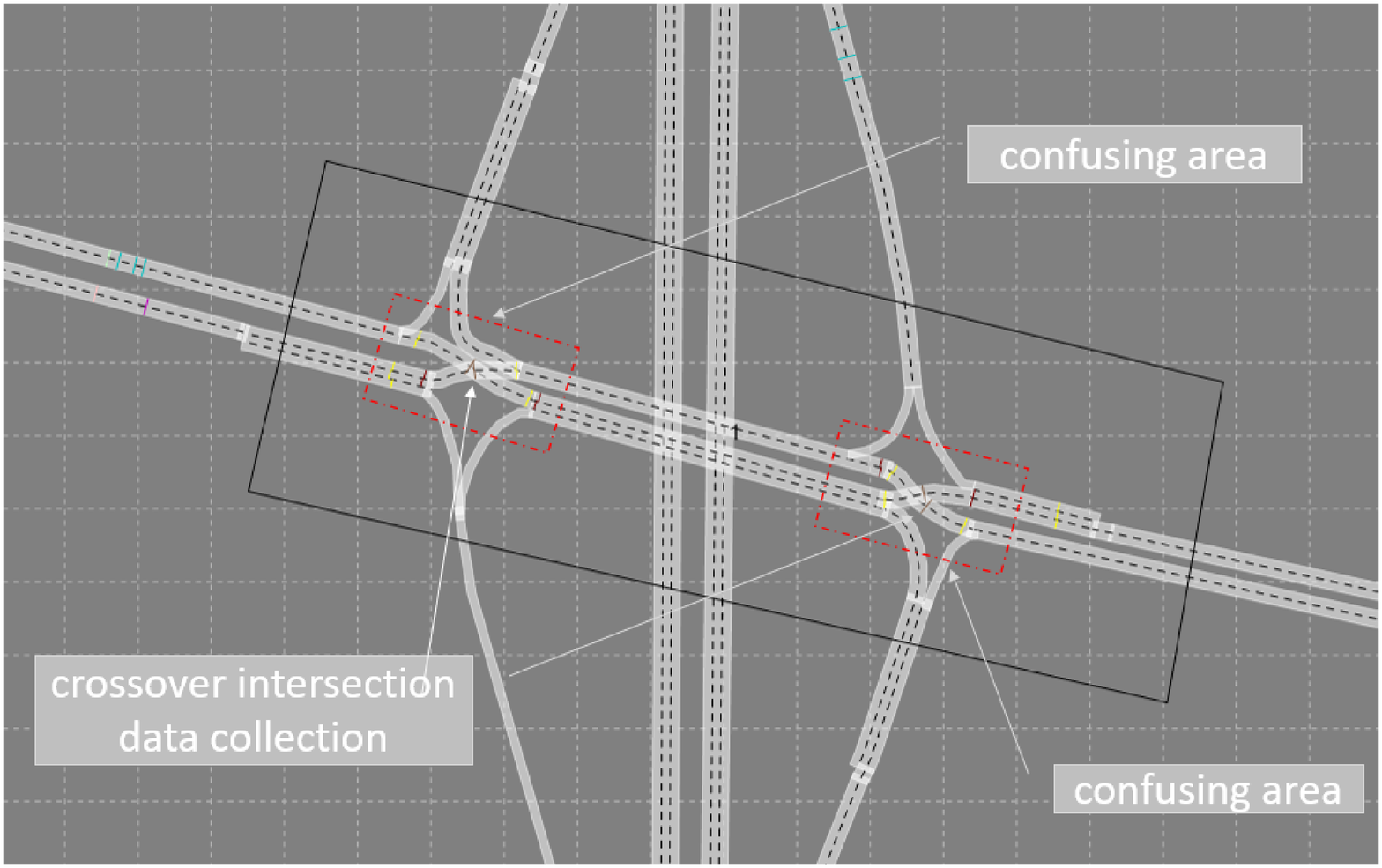}}
\end{minipage}\\
\begin{minipage}[h]{0.8\columnwidth}
\centering
\subfloat[RCUT Netowrk]{\includegraphics[scale=0.212]{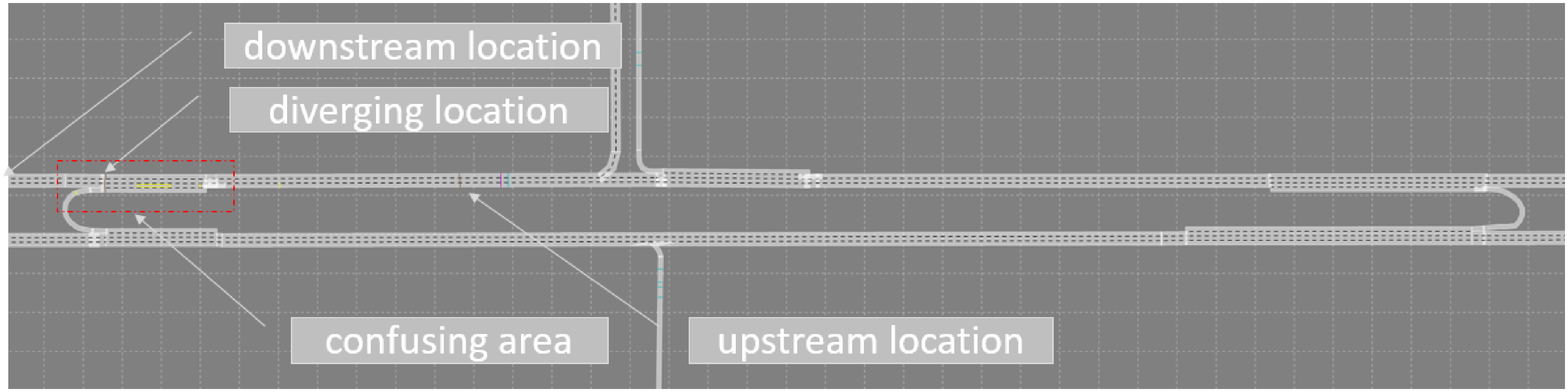}}
\end{minipage}
\caption{Configurations of selected DDI and RCUT }
\label{fig:geoUAID}
\end{figure}

The PTV Vissim, a microscopic traffic simulation, and its external driver model application programming interface (API) are used to develop the simulation network. We have constructed two AIDs:  a real-world DDI (Fig. \ref{fig:geoUAID}(a)) and a 1.61-mile, three-lane RCUT intersection Fig. \ref{fig:geoUAID}(b). 
The DDI is located at the intersection of the State Highway 72 (DE-72) and US Highway 13 (US-13).  It was converted from a convetional diamond interchange in early 2016 and open to trafic in late 2016 \cite{deDDI}. Four settings for DDI are simulated as shown in TABLE \ref{table:simCase}.

\begin{table}[h]
\centering
\caption{Simulation Cases for DDI}
\begin{tabular}{l|llll}
\hline
Case & CDI & DDI & AV & MPR \\ \hline
Base-CDI &  \checkmark &  &  & 0\% \\
Base-DDI &  &  \checkmark &  & 0\%\\
CAV-CDI & \checkmark  &  & \checkmark  & $10\text{-}100$\% \\
CAV-DDI &  & \checkmark  & \checkmark  & $10\text{-}100$\% \\ \hline
\end{tabular}
\label{table:simCase}
\end{table}

The arterial demand is assumed to be 3,000 vph for both westbound and eastbound direction. The traffic volume for either of the on-ramp is 400 vph. A CDI network is built for the comparison between a CDI and a DDI. Signalization is only implemented at the two cross-over locations in the DDI. Each through movements has a 55-s green phase in each signal cycle which is 120 s. For the CDI, the phase timings are set as 73 s, 17 s, and 18 s for through, left-turn to the on-ramp, and left-turn from the off-ramp, respectively. The speed limit is 50 mph for both of the networks. 
For the RCUT, only the westbound direction of the RCUT is analyzed. The distance between the minor street and the diverging point of the median U-turn is approximately 1,300 ft., larger than the 600-ft. minimal design requirement set forth by ASSHTO \cite{hancock2013policy} for RCUT. The mainline demand from the westbound direction is 5,000 vph and the demand from the southbound minor street is 400 vph.

For each level of MPR, ten replications of simulation is conducted to factor in the variability of the simulation. Each replication runs for 3,900 s with 300 s as the warm-up time to load the network with traffic. The simulation resolution is set as 10 Hz.  For studying the driver' confusion, 30 replications for each level of the confused drivers are conducted to obtain additional samples for the ANOVA test. The data collection is performed every 5 min.

%%%%%%%%%%%%%%%%%%%%%%%%%%%%%%%%%%%%%%%%%%%%%%%%%%%%%%%%%%%%%%%%%%%%%%%%%%%%%%%%%
\section{Results \& Discussion}
\label{sect:result}

\subsection{Impact of CAV}
\begin{figure} [h]
	\centering
	\frame{\includegraphics[width=0.9\columnwidth]{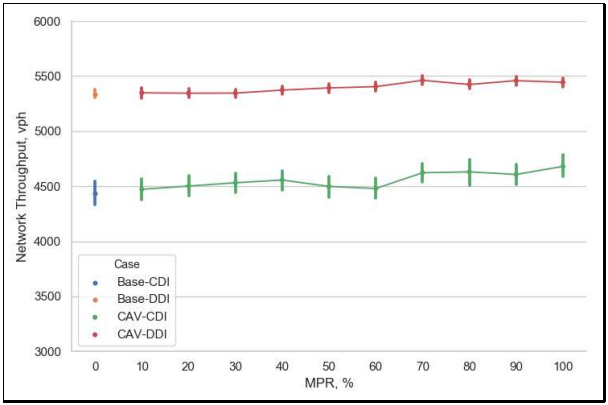}}   
	\caption{Network throughput} 
	\label{fig: netTPDDI}
\end{figure}

The network throughput of both the DDI and the CDI is shown in Fig. \ref{fig: netTPDDI}. The vertical bar associated with each marker represents the size of the 90\% confidence interval with bootstrapping \cite{haukoos2005advanced}, a statistical technique.  The throughput of the network increased to 5,350 vph in DDI from 4,400 vph that is observed in the CDI case. The standard deviation of the throughput in the CDI case is greater than that of the DDI.
With CAVs in the network, the overall trend for throughput for either DDI and CDI is increasing, given there are cases of slight deceases (i.e. 50\% and 60\% in the CDI case). Furthermore, with the same level of MPR, the observations in DDI exhibits a narrower 90\% confidence interval, an indication of less standard deviation.

\begin{figure} [h]
	\centering
	\frame{\includegraphics[width=0.9\columnwidth]{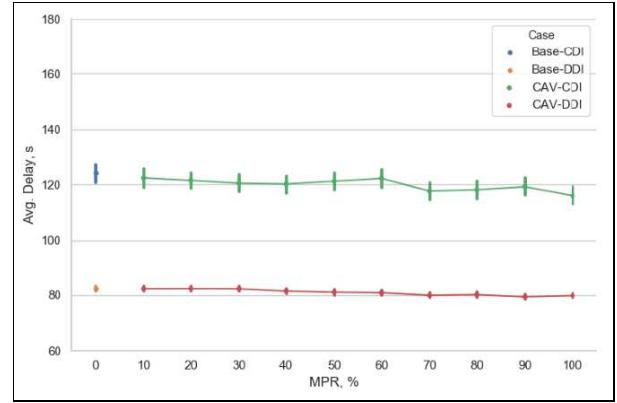}}   
	\caption{Average delay} 
	\label{fig: netAvgDelay}
\end{figure}

The average delay for each vehicle is plotted in Fig. \ref{fig: netAvgDelay}.  Similar to the throughput, the geometry configuration of the interchange greatly contributed to the reduction of the average delay. There is a clear separation (i.e. 40 s delay per vehicle) between the observations of DDI and those of the CDI. Again, the delay observed in DDI not only has a lower mean value, but also less standard deviation, compared to the CDI case. However, the delay only marginally decreases as the MPR increases. 

Both Fig. \ref{fig: netTPDDI} and Fig. \ref{fig: netAvgDelay} jointly indicate that only with the short-following distance and the zero start-up lost time do not significantly increase the performance of the signalized interchange.  
The start-up lost time is dedicated by the likelihood of a CAV being in the first vehicle at the stop line during a red phase. Even though zero start-up lost time are to be taken advantage of, the benefits from it would still be limited. For example for 120-s signal cycle within an hour, only 30 times per lane of such advantage is possible at best.
On the other hand, by reducing the signal phase and separating conflict, the network performance can be improved at a significant level. 
Therefore, AIDs could instead play more significant roles in improving the efficiency of a signalized intersection than CAV in terms of mobility.

%\begin{figure} [h]
%	\centering
%	\includegraphics[width=\columnwidth]{latentDemand.png}   
%	\caption{ Latent Demand} 
%	\label{fig: netLatentDemand}
%\end{figure}

\begin{figure} [h]
	\centering
	\includegraphics[width=\columnwidth]{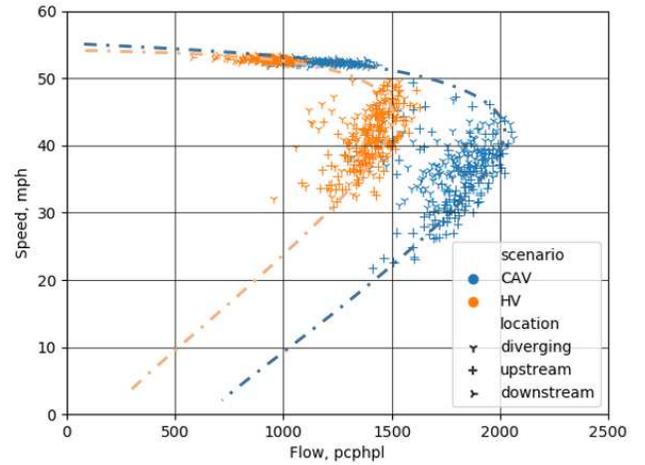}   
	\caption{Flow-speed curve observed at the diverging area for RCUT with full CAV penetration} 
	\label{fig: fsRCUT}
\end{figure}

When it comes to RCUT, the flow-speed observations in three locations (diverging, upstream, and downstream) are shown in Fig. \ref{fig: fsRCUT}.  In all three locations, the flow-speed curve of CAV systematically shifts to the higher flow rate region at the right side of the chart. The carrying capacity for the CAV case reaches 2,100 vph per lane. 
%However, the impact is localized as illustrated by the observation at the diverging and upstream location. 

\subsection{Impact of Driver's Confusion}
The corridor impact of the driver's confusion has not yet been taken into account in the previous studies. We consider the behaviors of drivers due to the confusion are: 1) sudden slowdown due to confusion prior to the AID ramp and 2) making an abrupt lane change at the last minute. The area for each AID that could most likely create confusion for drivers is identified in red in Fig. \ref{fig:geoUAID} based on the geometric design of the networks.  
In the RCUT, it is the U-turn pocket lane in the diverging area, which accommodates U- and left-turn traffic. The route decision point is set closer to the U-turn pocket lane to induce aggressive lane change that is likely observed from the unfamiliar drivers in order to make it to the U-turn lane. 
For the DDI, it is the signalized crossover intersections on the arterial. A reduction in desired speed is set for the unfamiliar drivers to mimic the slowdown behavior due to confusion. 

\begin{figure} [h]
	\centering
	\includegraphics[width=\columnwidth]{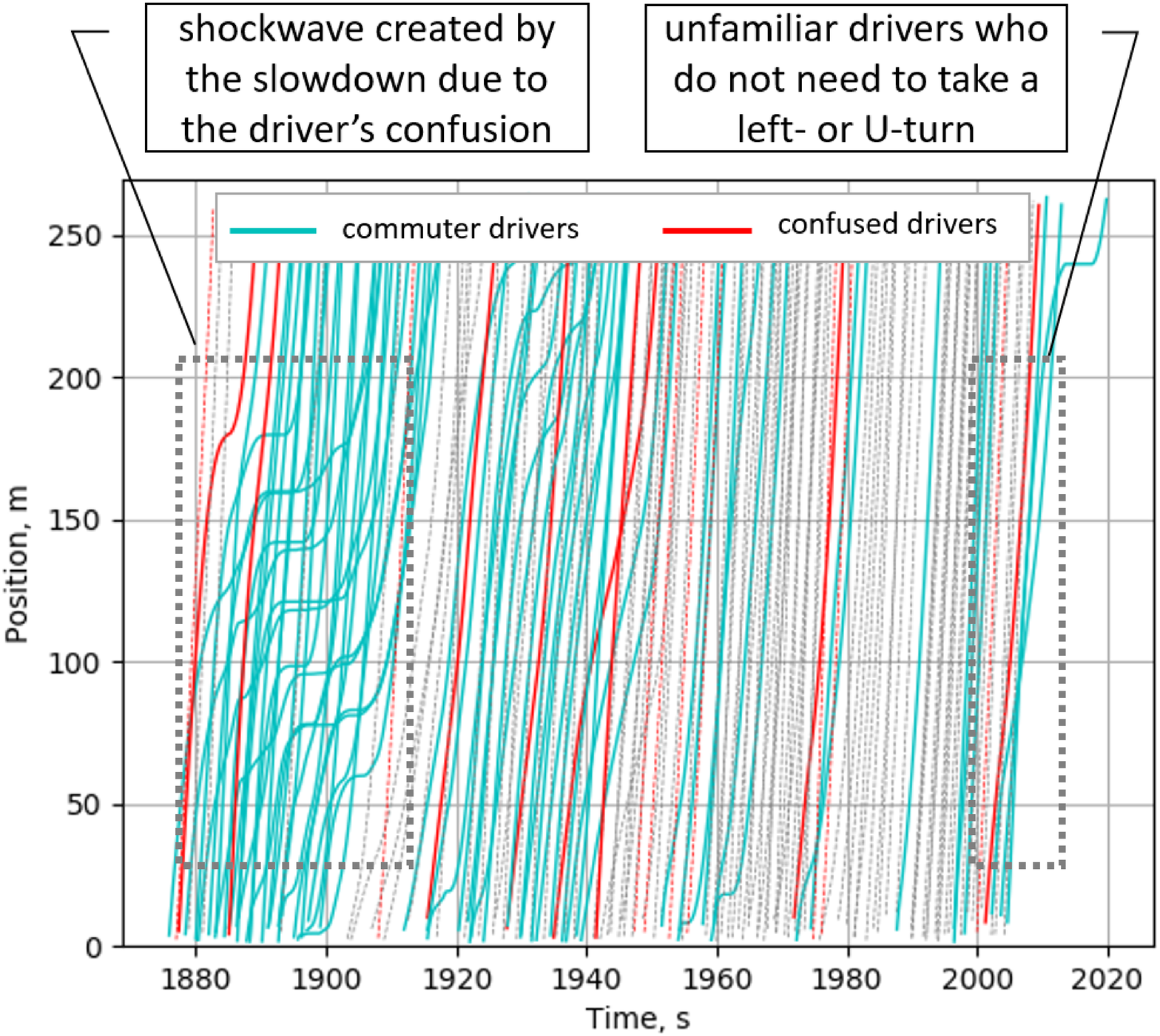}   
	\caption{Impact of driver's confusion} 
	\label{fig: impactEvaluationMethod}
\end{figure}

The percentage of unfamiliar drivers is set from 0\% to 20\% with a 5\% increment. For each scenario, 30 replications are run.  Point (road section) and network-wide performance data are collected every 5 min. The shockwave created by the driver’s confusion is illustrated in Fig. \ref{fig: impactEvaluationMethod}, where each line represents the trajectory of one vehicle from the simulation with 10\% unfamiliar drivers for RCUT. Red trajectory lines are unfamiliar drivers, whereas the cyan lines represent commuter drivers who are familiar and have gotten used to the RCUT. As seen, the sudden slowdown due to the driver’s confusion creates a shockwave and it propagates upstream, affecting the following vehicles.  On the right side of Fig. \ref{fig: impactEvaluationMethod}, the traffic trajectories indicate a free-flow condition in the absence of slowdown or abrupt lane change induced by the driver’s confusion.  As demonstrated, too much driver’s confusion could easily disrupt the progression of the traffic, not to mention the safety hazard it may create.

\begin{figure} [h]
	\centering
	\includegraphics[width=\columnwidth]{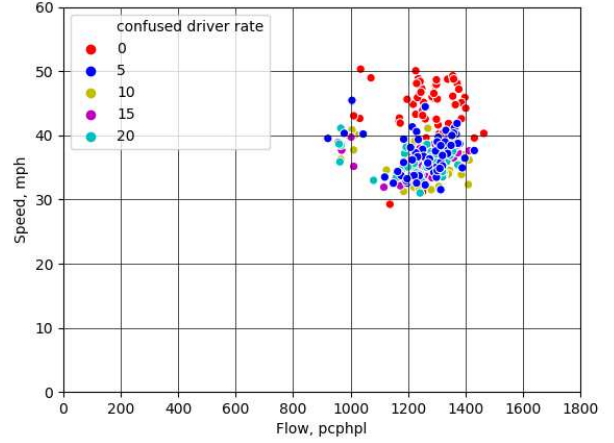}   
	\caption{Flow-speed curve observed at the diverging area for RCUT for driver’s confusion} 
	\label{fig: fsConfusion}
\end{figure}

The speed-flow diagram of the diverging area of the RCUT network is shown in Fig. \ref{fig: fsConfusion}. The overall speed of the traffic flow with confused drivers is lower than the base case. This is due to the temporary traffic obstruction of the unexpected behaviors of the confused drivers. The impacted vehicles at the end of the diverging area where the data are collected have not regained the prevailing speed of the roadway. As a result, the data sample points shift downward to the range of 30 mph and 40 mph with the presence of confused drivers.

The average vehicle delay for the entire network is collected. ANOVA test with post-hoc Tukey’s method \cite{pairwiseAnova} is conducted to assess the statistical difference among the five tested scenarios at 95\% confidence level. The ANOVA test result (TABLE \ref{table:TurkeyTestRCUT}) shows that the pairwise differences among five levels of confused drivers are statistically different. 
Similarly, the ANOVA test for average vehicle delay for DDI exhibits an increasing pattern that the average vehicle delays are statistically different at 95\% confidence level as shown in TABLE \ref{table:TurkeyTestDDI}.

\begin{table}[!]
\centering
\caption{ANOVA Test for Average Vehicle Delay in RCUT}
\begin{tabular}{llclllll}
\hline
\textbf{\begin{tabular}[c]{@{}l@{}}Confused \\ Driver Rate\end{tabular}} & \textbf{N} & \textbf{Delay, s/veh} & \multicolumn{5}{l}{\textbf{Grouping}} \\
\hline
\textbf{0\%} & 360 & 12.2 & A &  &  &  &  \\
\textbf{5\%} & 360 & 28.65 &  & B &  &  &  \\
\textbf{10\%} & 360 & 39.36 &  &  & C &  &  \\
\textbf{15\%} & 360 & 43.45 &  &  &  & D &  \\
\textbf{20\%} & 360 & 48.79 &  &  &  &  & E \\ \hline
\end{tabular}
\label{table:TurkeyTestRCUT}
\end{table}

\begin{table}[!]
\centering
\caption{ANOVA Test for Average Vehicle Delay in DDI}
\begin{tabular}{llclllll}
\hline
\textbf{\begin{tabular}[c]{@{}l@{}}Confused \\ Driver Rate\end{tabular}} & \textbf{N} & \textbf{Delay, s/veh} & \multicolumn{5}{l}{\textbf{Grouping}} \\
\hline
\textbf{0\%} & 360 & 81.42 & A &  &  &  &  \\
\textbf{5\%} & 360 & 82.44 &  & B &  &  &  \\
\textbf{10\%} & 360 & 83.54 &  &  & C &  &  \\
\textbf{15\%} & 360 & 84.41 &  &  &  & D &  \\
\textbf{20\%} & 360 & 85.78 &  &  &  &  & E \\ \hline
\end{tabular}
\label{table:TurkeyTestDDI}
\end{table}

%%%%%%%%%%%%%%%%%%%%%%%%%%%%%%%%%%%%%%%%%%%%%%%%%%%%%%%%%%%%%%%%%%%%%%%%%%%%%%%%%
\section{Conclusion}
\label{sect:conclusion}

The alternative intersection designs have attracted an increasing amount of attention as a promising measure to improve the performance of an intersection, as evidenced by field deployments and simulation study. The joint deployment of alternative intersection designs and CAV is studied in this paper via microscopic traffic simulation. According to the results on mobility, only 7\% increase in throughput is observed under full CAV market penetration, compared to the 20\% gain in throughput with only the conversion from a conventional diamond interchange to a diverging diamond interchange. Note that the benefits of the CAV could be further optimized in operation, such as using eco-driving approaching control, adaptive signal control, or ultimately with signal-free autonomous intersection management. They will be part of the future study.

The impact of the potential the driver's confusion is quantified by analyzing the traffic flow and vehicle trajectory data. It is found that the influence is more localized. Hence limited impact on performance at network level is observed. Future study should focus on the safety aspect at a more granular level (e.g., individual vehicle level).  Additionally, explicit consideration for the increased safety brought by CAV should be integrated into the subsequent study.
Lastly, more sophisticated scenarios, including signal plans, demand composition, CAV applications, etc., should be included to expand the comparison.

%\appendix
\appendices
\section{\textcolor{black}{List of Abbreviations}}
%\textcolor{red}{Definitions for the abbreviations used} \\ \newline
%\begin{table}[H]
%\centering
%\caption{\textcolor{red}{List of Abbreviations}} 
%\resizebox{0.9\textwidth}{!}
%{
\begin{tabular}{p{0.8in}|p{2.2in}} 
\hline  
\textbf{Abbreviation} & \textbf{Definition} \\ \hline 
AID & alternative intersection design \\ \hline
ANOVA & analysis of variance \\ \hline
API & application programming interface \\ \hline
DDI & diverging diamond interchange\\ \hline 
CDI & conventional diamond interchange \\ \hline 
RDT & roundabout \\ \hline
CAV & connected and automated vehicle \\ \hline
MUT & median U-turn intersection\\ \hline 
MPR & market penetration rate \\ \hline
DLT & displaced left-turn intersection\\ \hline 
RCUT & restricted crossing U-turn intersection  \\ \hline 
V2X & vehicle-to-anything\\ \hline 
V2I & vehicle-to-infrastructure \\ \hline
SAE & Society of Automotive Engineers \\ \hline
HV & human-driven vehicle\\ \hline 
MPR & market penetration rate\\ \hline 
ASSHTO & American Association of State Highway and Transportation Official\\ \hline  
%CDI & xxx\\ \hline 
%CDI & xxx\\ \hline  
\end{tabular}
%}
%\label{table:abbrv}
%\end{table}

\bibliographystyle{IEEEtran}
\bibliography{CAV_UAID_R1}
\end{document}